\overfullrule=0pt
\input harvmac
\def\a{{\alpha}}

\def\ah{{\widehat \a}}
\def\lh{{\widehat \lambda}}
\def\kh{{\widehat \kappa}}
\def\wh{{\widehat w}}

\def\l{{\lambda}}

\def\b{{\beta}}
\def\bh{{\widehat\beta}}

\def\g{{\gamma}}

\def\d{{\delta}}

\def\s{{\sigma}}
\def\k{{\kappa}}

\def\half{{1\over 2}}
\def\p{{\partial}}

\def\t{{\theta}}

\def\bar{\overline}

\Title{\vbox{\hbox{IFT-P.044/2004 }}}
{\vbox{
\centerline{\bf BRST Cohomology and Nonlocal Conserved Charges}}}
\bigskip\centerline{Nathan Berkovits\foot{e-mail: nberkovi@ift.unesp.br}}
\bigskip
\centerline{\it Instituto de F\'\i sica Te\'orica, Universidade Estadual
Paulista}
\centerline{\it Rua Pamplona 145, 01405-900, S\~ao Paulo, SP, Brasil}

\vskip .3in
A relation is found between 
nonlocal conserved charges in string
theory and certain ghost-number two states in the BRST cohomology.
This provides a simple proof that the nonlocal conserved charges for
the superstring in an $AdS_5\times S^5$ background are BRST-invariant
in the pure spinor formalism and are $\kappa$-symmetric
in the Green-Schwarz formalism.

\vskip .3in

\Date {September 2004}

\newsec{Introduction}

Maldacena's conjecture that $d=4$ ${\cal N}=4$ super-Yang-Mills theory
is dual to superstring theory in an $AdS_5\times S^5$ background
has been difficult to prove since the perturbative descriptions of
these two theories do not overlap. To obtain non-perturbative information
about the two theories, one possible tool could be integrability and
there have been various papers discussing this possibility both on
the super-Yang-Mills side and on the superstring side.

On the superstring side, Bena, Polchinski and Roiban \ref\bena
{I. Bena, J. Polchinski and R. Roiban, {\it Hidden Symmetries
of the $AdS_5\times S^5$ Superstring}, Phys. Rev. D69 (2004) 046002,
hep-th/0305116.} constructed an
infinite set of nonlocal conserved charges for the
Green-Schwarz (GS) superstring in an $AdS_5\times S^5$ background\foot
{These nonlocal charges were also independently found by Polyakov \ref
\polyakov{A.M. Polyakov, {\it Conformal Fixed Points of
Unidentified Gauge Theories}, Mod. Phys. Lett. A19 (2004) 1649,
hep-th/0405106.}. Similar nonlocal charges have been proposed
in \ref\mandal{G. Mandal, N.V. Suryanarayana and S.R. Wadia,
{\it Aspects of Semiclassical Strings in $AdS_5$}, Phys. Lett. B543 (2002) 81,
hep-th/0206103}.}, suggesting
an integrable structure. Vallilo \ref\vallilo{B.C. Vallilo,
{\it Flat Currents in the Classical $AdS_5\times S^5$ Pure Spinor
Superstring}, JHEP 0403 (2004) 037, hep-th/0307018.}
then constructed an analogous set of nonlocal
conserved charges using the pure spinor formalism for the
superstring in an $AdS_5\times S^5$ background \ref\pures{N. Berkovits,
{\it Super-Poincar\'e Covariant Quantization of the Superstring},
JHEP 0004 (2000) 018, hep-th/0001035\semi N. Berkovits and O. Chand\'{\i}a, 
{\it Superstring Vertex Operators in an $AdS_5\times S^5$ Background},
Nucl. Phys. B596 (2001) 185, hep-th/0009168\semi
B.C. Vallilo, {\it One-Loop Conformal Invariance of the Superstring
in an $AdS_5\times S^5$ Background}, JHEP 0212 (2002) 042, hep-th/0210064.}.
These nonlocal charges for the superstring were related in
\ref\dolan{L. Dolan,
C. Nappi and E. Witten, {\it A Relation Between Approaches to
Integrability in Superconformal Yang-Mills Theory}, JHEP 0310
(2003) 017, hep-th/0308089.}
to a corresponding
set of nonlocal charges on the
super-Yang-Mills side.

In the GS formalism for the superstring, 
invariance under $\kappa$-transformations is
crucial for determining the physical spectrum. For example,
classical $\k$-symmetry is preserved in a curved background when the
background satisfies the low-energy supergravity equations of motion, so
onshell
massless vertex operators must be $\k$-symmetric at least at the classical
level. Although quantum $\k$-transformations and massive GS vertex operators
are not yet understood, it is reasonable to expect that all physical GS
states should be $\k$-symmetric. This would imply that the conserved charges
for physical symmetries should also be $\k$-symmetric.

In the pure spinor formalism for the superstring,
the role of $\k$-symmetry is replaced by BRST invariance. In this case,
quantization and massive vertex operators are well-understood,
and the physical
spectrum is described by states in the BRST cohomology.
So physical symmetries in the pure spinor formalism must be BRST-invariant.

Surprisingly, it has not been previously
verified if the nonlocal conserved charges of \bena\ 
are $\k$-symmetric, or if the nonlocal conserved charges of \vallilo\
are BRST-invariant.
This led some people (including this author) to
conclude that the charges were not $\k$-symmetric and to
question their physical significance. Due to the
insistence of Witten \ref\witten{E. Witten, private communication.} that 
these nonlocal charges should describe physical symmetries in analogy with
the nonlocal charges on the super-Yang-Mills side \dolan,
the $\k$-symmetry and BRST
invariance of the nonlocal charges of 
\bena\ and \vallilo\ were investigated.

As described in section 2, 
the existence of an infinite set of
BRST-invariant nonlocal charges
can be deduced from the absence of certain states
in the BRST cohomology at ghost-number two. 
These ghost-number two states are $f^C_{AB} h^A h^B$ where
$h^A$ are the BRST-invariant ghost-number one states associated with
the global isometries and $f^C_{AB}$ are the structure constants.
Whenever $f^C_{AB} h^A h^B$ can be written as $Q \Omega^C$ for
some $\Omega^C$ (i.e. whenever $f^C_{AB} h^A h^B$ is not in the BRST
cohomology), one can construct an infinite set of BRST-invariant
nonlocal charges. It would be interesting to know if quantum corrections
to the ghost-number two cohomology are related to the potential anomalies
discussed in \ref\abdalla{E. Abdalla, M. Gomes and M. Forger,
{\it On the Origin of Anomalies in the Quantum Non-local Charge for the
Generalized Non-linear Sigma Model}, Nucl. Phys. B210 (1982) 181.}.
%Since quantum corrections are not expected to add or remove states
%from the BRST cohomology, this result suggests that if a 
%conformally invariant
%string theory has an infinite set of BRST-invariant nonlocal charges
%at the classical level, the string theory will also have an infinite set
%at the quantum level.

In section 3 of this paper, it is shown using the pure spinor formalism
for the
superstring in an
$AdS_5\times S^5$ background that the relevant 
ghost-number two states are absent from the
classical BRST cohomology.
The corresponding infinite set of BRST-invariant nonlocal
charges is then explicitly constructed and shown to coincide with
the conserved charges found by Vallilo in \vallilo.
It is also shown that the conserved charges of Bena, Polchinski and Roiban
in \bena\ are $\k$-symmetric.

\newsec{Relation of Nonlocal Charges with BRST Cohomology}

Suppose one has a BRST-invariant string theory with global physical
symmetries described by the charges $q^A = \int d\s ~j^A(\s)$.
Since these symmetries take physical states to physical states,
$q^A=\int d\s~j^A(\s)$ must satisfy $Q(q^A)=0$ where $Q$
is the BRST charge. Note that
$\{Q, b_0\}=H$ where $H$ is the Hamiltonian, so BRST invariance
implies charge conservation if $q^A$ commutes with the $b_0$
ghost, i.e. if $q^A$ can be chosen in Siegel gauge.
With the exception of the zero-momentum dilaton, it is expected that
all ghost-number zero states in the BRST cohomology
can be chosen in Siegel gauge.\foot{In the pure spinor formalism for
the superstring, there
is no natural $b$ ghost. Nevertheless, it is 
expected that for any ghost-number zero state in the pure-spinor BRST
cohomology, there exists a gauge in which the state is annihilated by $H$.}

Since $Q(\int d\s j^A(\s)) =0$,
$Q(j^A) = \p_\s h^A$
for some $h^A$ of ghost-number one. And $Q^2=0$ implies that 
$Q( \p_\s h^A)=0$, which implies that $Q(h^A)=0$ since
there are no $\s$-independent worldsheet fields.

Consider the BRST-invariant ghost-number two states $f^C_{AB}:h^A h^B:$
where $f^C_{AB}$ are the
structure constants
and normal-ordering is defined in any BRST-invariant manner,
e.g. $:h^A(z) h^B(z): \equiv {1\over{2\pi i}}\oint dy (y-z)^{-1}
h^A(y) h^B(z)$ where the contour of $y$ goes around the point $z$.
It will now be shown that whenever $f^C_{AB} :h^A h^B :$ is not in the BRST
cohomology\foot{This
BRST cohomology is defined in the ``extended'' Hilbert space which
includes the zero mode of the $x^m$ variables. As explained in \ref\belop
{A. Astashkevich and A. Belopolsky, {\it String Center of Mass Operator
and its Effect on BRST Cohomology}, Commun. Math. Phys. 186 (1997) 109,
hep-th/9511111.},
the inclusion of the $x^m$ zero mode in the Hilbert space 
allows global isometries
to be described by ghost-number one states in the cohomology.}, i.e. 
whenever there exists an operator
$\Omega^{C}$ satisfying $Q(\Omega^{C}) = f^C_{AB}:h^A h^B:$, one can
construct an infinite set of nonlocal BRST-invariant charges.

To prove this claim, consider the nonlocal charge
$$k^{C} = f^C_{AB} :\int_{-\infty}^\infty d\s ~j^A(\s)~ 
\int_{-\infty}^\s d\s' ~j^B(\s'): .$$
Using $Q( j^A) = \p_\s h^A$, one finds that 
$Q( k^{C}) = \int d\s l^{C}(\s)$ where
$$l^{C} = -2 f^C_{AB} :h^A(\s) 
j^B(\s):.$$
One can check that $Q(l^{C}) = f^C_{AB}\p_\s (:h^A h^B:)$, so
$Q(l^{C}-\p_\s\Omega^{C}) = 0$ where
$\Omega^{C}$ is the operator which is assumed to satisfy
$Q(\Omega^{C}) = f^C_{AB} : h^A h^B:$.

Since  
$(l^{C}-\p_\s\Omega^{C})$ has $+1$ conformal weight and since
BRST cohomology is only nontrivial at zero conformal weight, 
$l^{C}-\p_\s\Omega^{C} = Q(\Sigma^{C})$ for some $\Sigma^{C}$.
Using $\Sigma^{C}$, one can therefore construct the nonlocal
BRST-invariant charge 
$$\widetilde q^{C} =f^C_{AB} :\int_{-\infty}^\infty d\s ~j^A(\s)~ 
\int_{-\infty}^\s d\s' ~j^B(\s'): - 
\int_{-\infty}^{\infty} d\s \Sigma^{C}(\s) .$$

By repeatedly commuting $\widetilde q^{C}$ with 
$\widetilde q^{D}$, one generates
an infinite set of nonlocal BRST-invariant charges. So as claimed,
$f^C_{AB}:h^A h^B:=
Q(\Omega^{C})$ implies the existence of
an infinite set of nonlocal BRST-invariant charges.

\newsec{BRST-Invariant Charges in $AdS_5\times S^5$ Background}

The results of the previous section will now be applied
to the charges in
an $AdS_5\times S^5$ background using the pure spinor
formalism for the superstring.
As in the Metsaev-Tseytlin 
GS action in an $AdS_5\times S^5$ background
\ref\metsaev{R.R. Metsaev and A.A. Tseytlin
{\it Type IIB Superstring Action in $AdS_5\times S^5$ Background},
Nucl. Phys. B533 (1998) 109, hep-th/9805028.},
the action
in the pure spinor formalism \pures\ is constructed from left-invariant
currents $J^A = (g^{-1} \p g)^A$ where
$g(x,\t,\widehat\t)$ 
takes values in the coset $PSU(2,2|4)/SO(4,1)\times SO(5)$,
$A= ([ab],m,\a,\ah)$ ranges over the
30 bosonic and 32 fermionic elements in the Lie algebra of $PSU(2,2|4)$,
$[ab]$ labels the $SO(4,1)\times SO(5)$ ``Lorentz''
generators, $m=0$ to 9 labels
the ``translation'' generators, and $\a,\ah=1$ to 16 label the 
fermionic generators.
The action in the pure spinor formalism also involves left and
right-moving bosonic ghosts, $(\l^\a, w_\a)$ and $(\lh^\ah,\wh_\ah)$,
which satisfy the pure spinor constraints
$\l\g^m\l=\lh\g^m\lh=0$. These pure spinor ghosts transform
as spinors under the local $SO(4,1)\times SO(5)$ transformations
and couple to the
$AdS_5\times S^5$ spin connection in the worldsheet action through
their Lorentz currents
$N_{ab} = \half w\g_{ab}\l$ and $\widehat N_{ab} = \half \wh\g_{ab}\lh$.

To construct the nonlocal charges, the notation and conventions of \vallilo\ 
will
be used where
\eqn\notat{J_0 = (g^{-1}\p g)^{[ab]} T_{[ab]},\quad
J_1 = (g^{-1}\p g)^{\a} T_{\a},\quad
J_2 = (g^{-1}\p g)^{m} T_{m},}
$$
J_3 = (g^{-1}\p g)^{\ah} T_{\ah}, \quad N = \half (w\g^{[ab]}\l) T_{[ab]}$$
$$\overline J_0 = (g^{-1}\overline \p g)^{[ab]} T_{[ab]},\quad
\overline J_1 = (g^{-1}\overline \p g)^{\a} T_{\a},\quad
\overline J_2 = (g^{-1}\overline \p g)^{m} T_{m},$$
$$
\overline J_3 = (g^{-1}\overline \p g)^{\ah} T_{\ah},
\quad \widehat N = \half (\wh\g^{[ab]}\lh) T_{[ab]},$$
$$\p = \half ({\p\over{\p\tau}}
+{\p\over{\p\s}}), \quad
\overline\p = \half ({\p\over{\p\tau}}
-{\p\over{\p\s}}), $$
and $T_A$ are the $PSU(2,2|4)$ Lie algebra generators.
It will be convenient to also introduce the notation
$$\l = \l^\a T_\a,\quad \lh = \lh^\ah T_\ah.$$
Note that $\l$ and $\lh$ are fermionic since $(T_\a,T_\ah)$ are fermionic
and $(\l^\a,\lh^\ah)$ are bosonic.

Under classical
BRST transformations generated by 
$$Q = \int d\s (\l^\a J_3^\ah + 
\lh^\ah \bar J_1^\a) \d_{\a\ah} ,$$
$g$ transforms by right-multiplication as
\eqn\brst{Q (g) = g (\l+\lh)}
and the pure spinors transform as
$$Q (N) =  -2 [J_{3},\l],\quad Q( \widehat N )= 
-2  [\bar J_{1},\lh],\quad Q(\l)=Q(\lh)=0.$$
The left-invariant currents therefore transform as
$$Q (J_j) = \d_{j+3,0} \p\l + [J_{j+3},\l] +\d_{j+1,0}\p\lh + [J_{j+1},\lh],$$
$$Q (\bar J_j )=\d_{j+3,0}\bar\p\l + [\bar J_{j+3},\l] +
\d_{j+1,0}\bar\p\lh +
 [\bar J_{j+1},\lh],$$
where $j$ is defined modulo 4, i.e. $J_j \equiv J_{j+4}$.

To prove the existence of an infinite set of BRST-invariant charges,
one needs to find an $\Omega =\Omega^{C} T_C$ satisfying 
$Q\Omega = : \{ h, h\}:$ where $h= h^A T_A$, $Q (j) = \p_\s h$, and
$q^A= \int d\s j^A$ are the charges associated with the global
$PSU(2,2|4)$ isometries. It will be shown at the end of subsection (3.1) that
$Q (j) = \half \p_\s (g (\l -\lh) g^{-1}),$ so
\eqn\later{h = \half g (\l -\lh) g^{-1}.}

Note that $h$ is BRST-invariant since 
\eqn\brstinv{Q (h)=\half g \{(\l+\lh),(\l-\lh)\}g^{-1} =\half
g ((\l^\a\g^m_{\a\b}\l^\b) T_m -
(\lh^\ah\g^m_{\ah\bh}\lh^\bh) T_m) g^{-1} =0}
because of the pure spinor constraint.
Consider the ghost-number two state 
\eqn\considerh{\{h,h\} = \half
g (\l-\lh)(\l-\lh) g^{-1} = -\half g \{\l,\lh\} g^{-1}.}
Since $\{\l+\lh, \l+\lh\} = 2\{\l,\lh\}$, one can write this state as
$Q\Omega$ where
\eqn\omegae{\Omega = - {1\over 4} g(\l+\lh) g^{-1}.}
So $Q\Omega = \{h,h\}$, which implies
the existence of an infinite set of BRST-invariant charges. 

\subsec{Explicit construction of BRST-invariant nonlocal charges}

To explicitly construct these BRST-invariant charges, suppose one has
a current whose $\tau$-component $a$ satisfies
\eqn\asat{Q a = \p_\s \Lambda + [a,\Lambda]}
for
some $\Lambda$. Then the charge 
\eqn\patho{P (e^{-\int_{-\infty}^\infty d\s ~a(\s)}) \equiv}
$$
1-\int_{-\infty}^\infty d\s a (\s)+ 
\int_{-\infty}^\infty d\s a(\s)  
\int_{-\infty}^\sigma d\s' a(\s') -
\int_{-\infty}^\infty d\s a(\s)  
\int_{-\infty}^\sigma d\s' a(\s')    
\int_{-\infty}^{\sigma'} d\s'' a(\s'')  + ... $$
satisfies $Q(   
P (e^{-\int_{-\infty}^\infty d\s ~a(\s)}) )= 0$. So
$P (e^{-\int_{-\infty}^\infty d\s ~a(\s)})$ is a BRST-invariant charge.

To construct $a$ satisfying 
\asat, consider 
\eqn\anew{a(c_j,\bar c_j) = g (c_0 N + c_1 J_1 + c_2 J_2 + c_3 J_3
+
\bar c_0 \widehat N + \bar c_1 \bar J_1 + \bar c_2 \bar J_2 + 
\bar c_3 \bar J_3 ) g^{-1} }
where $c_j$ and $\bar c_j$ are constant coefficients.
Note that $a(c_j,\bar c_j)$ is invariant
under the local $SO(4,1)\times SO(5)$ transformations.

Using the BRST transformations of \brst, 
\eqn\dela{ Q a = g [\l+\lh, c_0 N + \bar c_0\widehat N
+ \sum_{k=1}^3( c_k J_k  + \bar c_k \bar J_k)] g^{-1}}
$$
+ g (-2 c_0 [J_3,\l] -2 \bar c_0 [J_1,\lh] +
c_1 \p\l + c_3 \p\lh + \bar c_1 \bar\p\l + \bar c_3\bar\p\lh ) g^{-1}
$$
$$+ g \sum_{k=1}^3 (c_k [J_{k+3},\l] + c_k [J_{k+1},\lh] +
\bar c_k [\bar J_{k+3},\l] + \bar c_k [\bar J_{k+1},\lh]) g^{-1} .$$
And defining 
$$\Lambda(b,\bar b) =  g (b\l + \bar b \lh) g^{-1} ,$$
where $b$ and $\bar b$ are constant coefficients, one obtains
\eqn\secondside{\p_\s \Lambda + [a,\Lambda] = 
g( b(\p\l-\bar\p\l) + \bar b (\p\lh-\bar\p\lh)) g^{-1} + 
g [\sum_{j=0}^3 (J_j - \bar J_j), b\l+\bar b \lh] g^{-1} }
$$
+ g [ c_0 N + \bar c_0\widehat N
+ \sum_{k=1}^3( c_k J_k  + \bar c_k \bar J_k), b\l+\bar b\lh ] g^{-1} .$$
Setting \dela\ equal to \secondside\ and using the worldsheet equations of
motion \pures\vallilo 
\eqn\eomone{\bar\p\l +[\bar J_0,\l] = -\half [\widehat N,\l],\quad
\p\lh +[J_0,\lh] = -\half [N,\lh],}
and the pure spinor constraints $[\l,N]=[\lh,\widehat N]=0$, one
obtains the conditions
\eqn\condone{c_1 =b, \quad -\bar c_3= \bar b ,}
$$-c_1 + c_2 =b c_1 +b,
\quad -c_2 + c_3 =b c_2 + b,\quad -c_3 -2 c_0 = b c_3 +b,$$
$$-c_1 = \bar b c_1 + \bar b,\quad -c_2 + c_1 = \bar b c_2 + \bar b, \quad
-c_3 + c_2 = \bar b c_3+\bar b, \quad 2 c_0 + c_3 = -2 \bar b c_0 + \bar b,$$
$$ 2\bar c_0 + \bar c_1= -2b\bar c_0 -b,\quad
-\bar c_1 + \bar c_2 =b \bar c_1 -b,\quad -\bar c_2 +\bar  c_3 
=b\bar  c_2 - b,\quad -\bar c_3 = b\bar c_3 -b,$$
$$-\bar c_1 -2\bar c_0 = \bar b \bar c_1 - \bar b,\quad -\bar c_2 + \bar c_1 
= \bar b \bar c_2 - \bar b, \quad
-\bar c_3 +\bar  c_2 = \bar b \bar c_3-\bar b.$$

The conditions of \condone\ are solved by 
\eqn\soluone{c_0 = \half(1- \mu^2 ),\quad  c_1 = \pm \mu^{1\over 2} -1,
\quad c_2 = \mu -1, \quad c_3 = \pm\mu^{3\over 2} -1, } 
$$\bar c_0 =\half( \mu^{-2}-1), \quad \bar c_1 = 1\mp\mu^{-{3\over 2}},\quad
\bar c_2 = 1-\mu^{-1},\quad \bar c_3 =1\mp \mu^{-\half}  ,$$
$$b = \pm \mu^\half -1, \quad \bar b = \pm\mu^{-\half} -1,$$
which coincides with the solution of \vallilo\ for conserved currents.

Note that the global charge $q=\int d\s j(\s)$ can be
obtained from \anew\ by expanding $a(\mu)$ near $\mu=1$. If $\mu=1+\epsilon$,
one finds that $a(\mu)= \epsilon j + 
{\cal O}(\epsilon^2)$ and $\Lambda (\mu) = \epsilon h + 
{\cal O}(\epsilon^2)$ where $Q (j) = h$.
So from the formulas 
$$b(\mu) = \mu^\half -1 = \half \epsilon +{\cal O}(\epsilon^2),\quad
\bar b(\mu) = \mu^{-\half} -1 = -\half \epsilon +{\cal O}(\epsilon^2),$$
one learns that 
\eqn\hexp{h = \lim_{\epsilon \to 0}
\epsilon^{-1}\Lambda (\mu) =\half g (\l-\lh) g^{-1},}
as was claimed in \later.

\subsec{$\k$-symmetry of nonlocal GS charges}

Finally, it will be shown
that the nonlocal GS conserved charges
of \bena\ are $\k$-symmetric.
In conformal gauge for the GS superstring,
the $\k$-transformations of $g$ and the $\k$-transformations
of the left-invariant currents
can be obtained from the BRST transformations of \brst\ by replacing
$\l$ with $[\k, J_2]$ and replacing $\lh$ with $[\kh, \bar J_2]$ where
$\k= \k^\ah T_\ah$ and $\kh = \kh^\a T_\a$.
This is the $AdS_5\times S^5$ version of the procedure adopted in \ref\tonin
{I. Oda and M. Tonin, {\it On the Berkovits Covariant Quantization of
GS Superstring}, Phys. Lett. B520 (2001) 398, hep-th/0109051\semi
N. Berkovits, {\it Towards Covariant Quantization of the Supermembrane},
JHEP 0209 (2002) 051, hep-th/0201151.}
where $\l^\a$ is replaced by $\Pi^m (\g_m\k)^\a$ and 
$\lh^\ah$ is replaced by $\bar\Pi^m (\g_m\kh)^\ah$. In GS
conformal gauge, $\eta_{mn} J_2^m J_2^n = \eta_{mn}\bar J_2^m \bar J_2^n=0$ 
and the
$\k$-transformations are constrained to satisfy $\k^\ah \bar J_1^\a 
\d_{\a\ah} =
\kh^\a J_3^\ah \d_{\a\ah}=0$ 
so that the $h_{zz}$ and $h_{\bar z\bar z}$ components
of the worldsheet metric do not transform. Together with the GS equations
of motion $[J_2,\bar J_1] =0$ and $[\bar J_2, J_3]=0$, these conditions
imply that 
\eqn\implyk{[\l, J_2] = [\lh, \bar J_2] = [\l,\bar J_1 ] = [\lh, J_3]=0.}

Using the current of \anew\ with $c_0=\bar c_0=0$, one finds that 
\dela\ is equal to \secondside\ if
\eqn\condtwo{c_1=b,\quad -\bar c_3 =\bar b,}
$$-c_1 + c_2 =b c_1 +b,
\quad -c_3 = b c_3 +b,$$
$$-c_1 = \bar b c_1 + \bar b,\quad -c_2 + c_1 = \bar b c_2 + \bar b,
\quad c_3 = \bar b, $$
$$\bar c_1 = - b,\quad -\bar c_2 +\bar  c_3 
=b\bar  c_2 - b,\quad -\bar c_3 = b\bar c_3 -b,$$
$$-\bar c_1  = \bar b \bar c_1 - \bar b,\quad  
-\bar c_3 +\bar  c_2 = \bar b \bar c_3-\bar b.$$

The conditions of \condtwo\ are solved by 
\eqn\solutwo{c_1 =  \pm \mu^{1\over 2} -1,
\quad c_2 = \mu -1, \quad c_3 = \pm\mu^{-\half} -1, } 
$$\bar c_1 = 1\mp \mu^\half, \quad \bar c_2 = 1-\mu^{-1},\quad
\bar c_3 = 1\mp\mu^{-\half},$$
$$b = \pm \mu^\half -1, \quad \bar b = \pm\mu^{-\half} -1,$$
which coincides with the conserved GS charges of \bena.

\vskip 15pt
{\bf Acknowledgements:} I would especially like to thank  
Edward Witten for stressing that
the nonlocal charges in an $AdS_5\times S^5$ background
should be $\kappa$-symmetric. I would also like to thank
Radu Roiban, Brenno Carlini Vallilo
and Edward Witten for useful discussions and for reading the draft,
CNPq grant 300256/94-9, 
Pronex 66.2002/1998-9,
and FAPESP grant 99/12763-0
for partial financial support, and the Simons Workshop in Mathematics
and Physics at SUNY at Stony Brook
for their hospitality.

\listrefs

\end